\documentclass[lettersize,journal]{IEEEtran}
\IEEEoverridecommandlockouts
\usepackage{amsmath,amssymb,amsfonts}
\usepackage{algorithm, algorithmic}
\usepackage{graphicx}
\usepackage{textcomp}
\usepackage{xcolor}
\usepackage{cite}
\usepackage{amsmath,amssymb,amsfonts}
\usepackage{graphicx, subcaption}
\usepackage{balance}
\usepackage{textcomp}
\usepackage{xcolor}
\usepackage{commath}
\usepackage{graphicx, subcaption}
\usepackage{booktabs}
\usepackage{dblfloatfix}
\usepackage{soul}
\renewcommand{\baselinestretch}{0.916}

\def\BibTeX{{\rm B\kern-.05em{\sc i\kern-.025em b}\kern-.08em
     T\kern-.1667em\lower.7ex\hbox{E}\kern-.125emX}}

\begin{document}

\makeatletter
\def\maketitle{%
  \par
  \begingroup
  \normalfont
  \renewcommand{\@makefnmark}{}
  \@maketitle
  \@thanks
  \endgroup
  \setcounter{footnote}{0}%
  \let\maketitle\relax
  \let\thanks\relax
}
\makeatother

\title{\large \textbf{RIS-Assisted Physical Layer Security: Artificial Noise-Driven Optimization and Measurements} }
\author{\small Ahmet Muaz Aktas,  
        Sefa Kayraklik, \IEEEmembership{\small Graduate Student Member,~IEEE,}
		Sultangali Arzykulov, \IEEEmembership{\small Senior Member,~IEEE,}
        Galymzhan Nauryzbayev, \IEEEmembership{\small Senior Member,~IEEE,}
        Ibrahim Hokelek, \IEEEmembership{\small Member,~IEEE,}
        Ali Gorcin, \IEEEmembership{\small Senior Member,~IEEE}
        \thanks{Copyright (c) 2026 IEEE. Personal use of this material is permitted. However, permission to use this material for any other purposes must be obtained from the IEEE by sending a request to pubs-permissions@ieee.org.}
        \thanks{A. M. Aktas, S. Kayraklik, I. Hokelek, A. Gorcin are with the Communications and Signal Processing Research (HİSAR) Lab., TÜBİTAK-BİLGEM, Kocaeli, Turkiye.  E-mail: \{muaz.aktas, sefa.kayraklik, ibrahim.hokelek\}@tubitak.gov.tr. S. Kayraklik is also with the Department of Electrical and Electronics Engineering, Koc University, Sariyer, Istanbul, Turkiye. S. Arzykulov and G. Nauryzbayev are with the Department of Electrical and Computer Engineering, School of Engineering and Digital Sciences, Nazarbayev University, Astana, Kazakhstan. Email: \{sultangali.arzykulov, galymzhan.nauryzbayev\}@nu.edu.kz. A. M. Aktas and A. Gorcin are also with the Electronics and Communication Department, Istanbul Technical University, Istanbul, Turkiye. E-mail: aligorcin@itu.edu.tr}
}
\maketitle

\begin{abstract}
Reconfigurable intelligent surface (RIS) has emerged as a key enabler for providing signal coverage, energy efficiency, reliable communication, and physical layer security (PLS) in next-generation wireless communication networks. This paper investigates an artificial noise (AN)-driven RIS-assisted secure communication system. The RIS is partitioned into two segments, where the first segment is configured to direct the communication signal (CS) toward the legitimate user (Bob), and the other one is configured to steer the AN toward the eavesdropper (Eve). To this end, iterative and discrete Fourier transform-based algorithms are developed for practical RIS phase shift optimization. The power allocation between the CS and the AN signals is optimized in such a way that the secrecy capacity (SC) is maximized while limiting Eve's channel capacity. The proposed PLS framework is evaluated through both simulations and software defined radio based testbed experiments. The results demonstrate promising improvements in the SC, highlighting the potential of AN-driven RIS-assisted PLS for practical deployments.
\end{abstract}

%\vspace{-8pt}

\section{Introduction}
With the rapid advancement of wireless communication technologies driven by mission-critical applications such as autonomous vehicles, industrial automation, public safety, and emergency services, ensuring transmission availability, confidentiality, and integrity has become a matter of paramount importance \cite{10409564}. Given the computational and energy overhead associated with upper-layer security mechanisms, physical-layer security (PLS) becomes a viable approach for ensuring secure transmission in latency-sensitive and resource-constrained environments. Reconfigurable intelligent surfaces (RISs) have emerged as an effective tool for PLS since the wireless propagation environment can be controlled by intelligently adjusting the phase shifts of their reflecting antenna elements. RISs can enhance signal quality at legitimate receivers while simultaneously degrading the channels of potential eavesdroppers, thereby improving both reliability and security. Artificial noise (AN) can be directed toward the eavesdroppers with the support of RISs to further degrade the channels of eavesdroppers through structured interference \cite{10736549}.

The number of studies that apply RISs for PLS has been rapidly increasing \cite{8723525, 9134962,10068724,10200914,10423876,kompostiotis2025optimizing}. For example, in \cite{8723525}, joint optimization of transmit beamforming and RIS phase shifts maximizes the secrecy capacity (SC). The secrecy outage probabilities of the RIS-assisted wireless communication with direct transmitter–receiver links are analyzed in \cite{9134962}. A low-complexity beamforming scheme that maximizes the SC while minimizing energy consumption is proposed in \cite{10068724}. The experimental studies show that tuning RIS phase shifts strengthens the legitimate link while weakening the eavesdropper's link \cite{10200914, 10423876}. Furthermore, \cite{kompostiotis2025optimizing} presents and experimentally validates a novel RIS phase-shift optimization method for multipath environments in RIS-assisted PLS systems. \textcolor{black}{ To account for more realistic network environments, \cite{9439833} investigates the enhancement of PLS in RIS-aided networks under dynamic eavesdropper locations. Building upon this, the secrecy performance of RIS-assisted systems over Rician fading channels in the presence of spatially random eavesdroppers is analyzed in \cite{10387228}. Furthermore, \cite{10315044} employs stochastic geometry to evaluate the SC for randomly located eavesdroppers in simultaneously transmitting and reflecting RIS networks.}

Recent studies further enhance PLS in RIS-assisted systems using AN-based transmission \cite{8972400,9133130,9201173,10066528,10901057}. The authors in \cite{8972400} formulate a secrecy rate optimization problem for an AN-enabled RIS-PLS system and show that AN is essential in challenging settings such as the presence of multiple eavesdroppers. In \cite{9133130}, the authors examine a multi-RIS, multi-user, MISO system with multi-antenna eavesdroppers, jointly optimizing RIS phases, transmit beamforming, and AN covariance to maximize the sum rate under a signal leakage constraint. In \cite{9201173}, the secrecy rate maximization for AN-based RIS-aided MIMO is studied through block coordinate descent and majorization–minimization methods. The authors in \cite{10066528} derive closed-form solutions for RIS partitioning with AN to maximize the SC under power constraints. Finally, \cite{10901057} presents an experimental testbed for validating a deep reinforcement learning to improve the SC in an AN-based RIS-PLS system. However, their solution is based on intelligent beam pair selection for the RIS partitions, as the communication and AN signals are transmitted from a single source. {\color{black} Furthermore, theoretical studies such as \cite{8972400,9133130,9201173,10066528} often rely on the assumption of perfect channel state information (CSI) and computationally intensive optimization techniques. 

Motivated by the gap between theoretical limits and practical deployment, this paper proposes an alternative architecture: an AN-based, RIS-assisted PLS scheme in which the communication signal (CS) and AN are physically decoupled at the transmitter. By steering these distinct signals to dedicated RIS partitions, our design achieves high spatial isolation and simplifies RIS phase shift control. The phase shifts of the first partition are optimized to enhance the received signal power at the legitimate user, whereas the second partition is configured to maximize interference at the eavesdropper. Furthermore, unlike prior works requiring perfect CSI or continuous phase shifts, our system utilizes practical, low-complexity iterative and discrete Fourier transform (DFT)-based codebook algorithms that rely solely on received signal power measurements to configure discrete binary phase shifts. Finally, we provide extensive over-the-air validation using a sub-6 GHz software-defined radio (SDR) testbed, demonstrating the system's viability under real-world hardware constraints. The main contributions can be summarized as follows:}
\begin{figure*}[t]
    \centerline{\includegraphics[width = 1\linewidth]{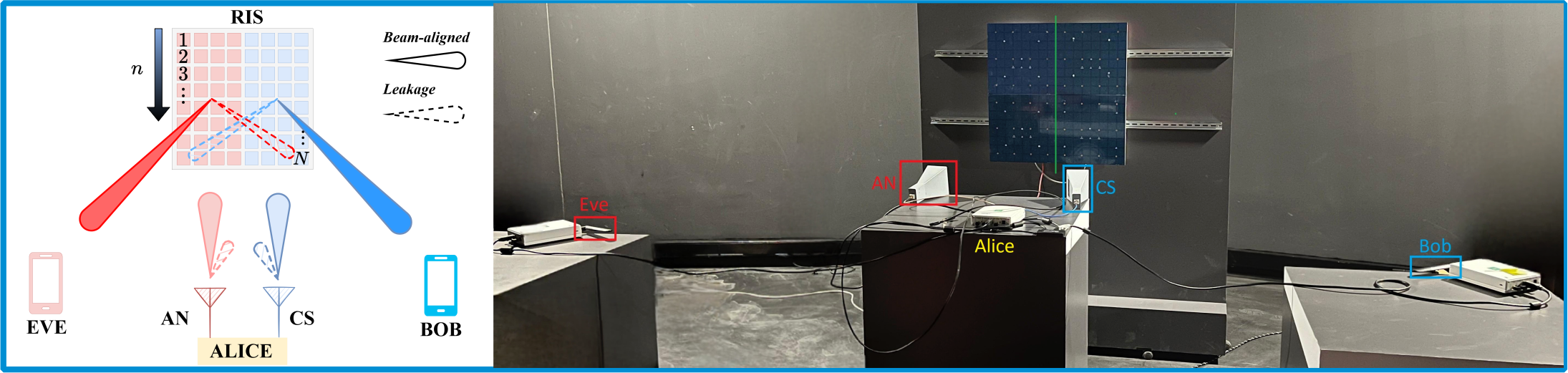}}
        \caption{AN-based RIS-assisted PLS system (left) and an illustration of the experimental setup (right).}
        \label{fig:measurementsetup}\vspace{-10pt}
\end{figure*}

\begin{itemize}
\item We design and implement a sub‑$6$ GHz RIS‑assisted PLS system in which a $256$-element RIS is partitioned to steer the CS to Bob and the AN signal to Eve. The design is validated in MATLAB and on an SDR based testbed with dual horn antennas. 
\item We derive Bob/Eve capacities and formulate the SC maximization with an explicit leakage control, guiding the CS/AN power allocation.
\item  We implement iterative and DFT-based algorithms for practical binary phase shift optimization of RIS antenna elements. Within $256$ trials, the iterative method increases the SC monotonically and converges to a higher SC value compared to the DFT-based method.
\item Optimization analysis shows that limiting Eve's capacity to $1\%$ of Bob's capacity requires more power allocation to AN, and hence lowers the SC, quantifying the power–secrecy trade-off for deployment planning.
\item The proof-of-concept testbed experiments, which are validated through extensive simulations, confirm the feasibility and effectiveness of the proposed PLS system in achieving notable SC improvements.
\end{itemize}

\section{System Model}
The system model is illustrated in Fig. \ref{fig:measurementsetup}, where PLS is realized through the RIS and the AN transmission signal. A transmitter (Alice) is equipped with two single-antenna elements, one of which is dedicated to transmitting the CS, denoted by $m_{cs}$, and the other is utilized to generate the AN signal, denoted by $m_{an}$, where $\mathbb{E} |m_{cs}|^2 = 1$ and $\mathbb{E} |m_{an}|^2 = 1$. There are two receivers representing the legitimate user (Bob) and the eavesdropper (Eve). {\color{black} To establish the theoretical and experimental baseline for the proposed architecture, we adopt the common assumption that Eve is an active but untrusted node in the network, meaning that the transmitter can estimate Eve's location or approximate channel state to steer the AN.} The RIS, which is partitioned vertically into two segments, is utilized to control the wireless propagation characteristics of the CS and the AN signals such that the right segment of the RIS is configured to strengthen the received signal powers of the CS at Bob, while the left segment is configured to direct the AN signal toward Eve. The RIS is composed of $N$ reflecting elements, where $N/2$ elements are allocated for Bob, and the remaining $N/2$ elements for Eve.\footnote{\color{black} In this correspondence, an equal partition is adopted to provide a robust baseline that aligns with the physical symmetry of the hardware testbed and the practical beamwidths of the transmit antennas. Preliminary numerical evaluations indicate that dynamically adjusting the partition size can yield further secrecy capacity gains by finely balancing the CS enhancement against AN interference; however, an exhaustive exploration of this dynamic trade-off is left for future work.}
In addition, the total transmit power of Alice is represented by $P_t = \alpha_1 P_{t} + \alpha_2 P_{t}$, where $\alpha_1 P_{t}$ and $\alpha_2 P_{t}$ represent the transmit power values of the CS and AN signals, respectively. The coefficients $\alpha_1$ and $\alpha_2$ denote the power scaling factors applied to the CS and AN signals, respectively, where $\alpha_1 + \alpha_2 = 1$. Since there is no line-of-sight between the transmitter and receivers, only the reflected signals from the RIS reach the receivers. The channel between Alice and the $n$-th RIS element is defined as
\begin{equation}
h_{pn} = |h_{pn}|e^{-j\phi_{pn}}, \quad \forall p \in \{s,a\},
\end{equation}
where $|h_{pn}|$ and $\phi_{pn}$ denote the amplitude and phase of the channel, respectively. The indices $s$ and $a$ are associated with the CS and AN transmitters, respectively. The channels from the $n$-th RIS element to Bob and Eve are given by
\begin{equation}
h_{nu} = |h_{nu}|e^{-j\phi_{nu}}, \quad \forall u \in \{b,e\},
\end{equation}
where $|h_{nu}|$ and $\phi_{nu}$ refer to the amplitude and phase components of the channel. The indices $b$ and $e$ represent Bob and Eve, respectively. Furthermore, the cascaded RIS channel representing the end-to-end link from the CS transmitter of Alice to Bob and Eve is expressed as
{\small
\begin{equation}
G_{sr_i l} = 
\begin{cases}
\quad\sum\limits_{n = 1}^{N/2} \quad h_{sn} e^{-j \theta_{n}} h_{nu}, & \text{if } i = e, \\ \\
\sum\limits_{n = N/2 + 1}^{N} h_{sn} e^{-j \theta_{n}} h_{nu}, & \text{if } i = b,
\end{cases}
\quad \forall l \in \{b,e\},
\end{equation}
}where $r_i$ identifies the selected partition of RIS, with $r_b$ and $r_e$ associated with Bob and Eve, respectively. The RIS partition dedicated to \textcolor{black}{Eve} consists of $N/2$ elements indexed by $n=1,2,\ldots,N/2$. Here, $\theta_{n}$ denotes the adjustable phase induced by the $n$-th reflecting element of RIS. The cascaded channel from the AN transmitter of Alice to Bob and Eve is given by
{\small
\begin{equation}
G_{ar_i l} = 
\begin{cases}
\quad\sum\limits_{n = 1}^{N/2} \quad h_{an} e^{-j \theta_{n}} h_{nu}, & \text{if } i = e, \\ \\
\sum\limits_{n = N/2 + 1}^{N} h_{an} e^{-j \theta_{n}} h_{nu}, & \text{if } i = b,
\end{cases}
\quad \forall l \in \{b,e\},
\end{equation}
}where the RIS partition associated with \textcolor{black}{Bob} consists of $N/2$ elements indexed by $n=N/2+1,N/2+2,\ldots,N$. 

\begin{figure*}[t]
\begin{equation}
    \begin{aligned}
        y_{bob}  =& \underbrace{\sqrt{P_{sr_bb}}G_{sr_bb}m_{cs}}_{\beta_1\, \text{ (aligned CS)}} 
        \, + \, \underbrace{\sqrt{P_{sr_eb}}G_{sr_eb}m_{cs}}_{\beta_2\, \text{(non-aligned CS)}} \, + \, \underbrace{\sqrt{P_{ar_eb}}G_{ar_eb}m_{an}}_{\beta_3\, \text{(non-aligned AN)}}
        \, + \, \underbrace{\sqrt{P_{ar_bb}}G_{ar_bb}m_{an}}_{\beta_4\, \text{(non-aligned AN)}} \, + \, 
        n_{w_b}
        \label{eq:y_bob_rec}
    \end{aligned}
\end{equation}

\begin{equation}
    \begin{aligned}
        y_{eve}  =& \underbrace{\sqrt{P_{ar_ee}}G_{ar_ee}m_{an}}_{\beta_5\, \text{(aligned AN)}} 
        \, + \, \underbrace{\sqrt{P_{ar_be}}G_{ar_be}m_{an}}_{\beta_6\, \text{(non-aligned AN)}}
        \, + \,  \underbrace{\sqrt{P_{sr_be}}\, G_{sr_be}m_{cs}}_{\beta_7\,\text{(non-aligned CS)}}
        \, + \, \underbrace{\sqrt{P_{sr_ee}}G_{sr_ee}m_{cs}}_{\beta_8\,\text{(non-aligned CS)}} \, + \, 
        n_{w_e}
        \label{eq:y_eve_rec}
    \end{aligned}
\end{equation}
\hrule
\end{figure*}

The received signal at Bob is given in (\ref{eq:y_bob_rec}) on the next page, where $\beta_1$, $\beta_2$, $\beta_3$, and $\beta_4$ correspond to the aligned CS, non-aligned CS, and the first and second non-aligned AN components, respectively. The term $n_{w_b}$ denotes the additive white Gaussian noise (AWGN) at Bob. Similarly, the received signal at Eve is expressed in (\ref{eq:y_eve_rec}), where $\beta_5$, $\beta_6$, $\beta_7$, and $\beta_8$ are associated with the aligned AN, non-aligned AN, and the first and second non-aligned CS components, respectively. Furthermore, applying the power scaling factor $\alpha_1$ to the CS, the received signal power at $u \,\, (\forall u \in \{b,e\})$ becomes $P_{sr_iu} = \alpha_1 P_{t} \times L_{sr_i}\times L_{r_iu}$ with the RIS partitions $(\forall r_i \in \{r_b,r_e\})$, where $L_{sr_i}$ and $L_{r_iu}$ represent the path losses between the CS transmitter and the RIS, and between the RIS and the users, respectively. Applying the power scaling factor $\alpha_2$ to the AN, the received signal power at $u \,\, (\forall u \in \{b,e\})$ becomes $P_{ar_iu} = \alpha_2 P_{t} \times L_{ar_i}\times L_{r_iu},$ where $L_{ar_i}$ and $L_{r_iu}$ refer to the path losses between the AN transmitter and the RIS, and between the RIS and users, respectively, while $n_{w_e}$ denotes the AWGN at Eve. 

\section{AN-based RIS-assisted PLS}
This section derives the SC of the AN-based RIS-assisted PLS framework using the received signal powers at Bob and Eve. It then analyzes the proposed system and develops practical optimization strategies for RIS with discrete phase shifts.

\subsection{SC Derivation}
To evaluate the SC, the channel capacities at Bob and Eve are first determined. The capacity of Bob, as a function of the transmit power ratio $\alpha$, can be derived sequentially as
\begin{align}
\label{eq:BobCapacity}
   C_b  &= \log_2 \hspace{-0.1cm} \left( \hspace{-0.1cm} 1 \hspace{-0.1cm} + \hspace{-0.1cm} \frac{\beta_1^2 + \beta_2^2}{\beta_3^2 + \beta_4^2+\sigma^2_{w_b}}\right)
     =\log_2 \hspace{-0.1cm} \left(\frac{|y_{bob}|^2}{\beta_3^2 + \beta_4^2+\sigma^2_{w_b}}\right),
\end{align}
where $|y_{bob}|^2 = \beta_1^2 + \beta_2^2 + \beta_3^2 + \beta_4^2+\sigma^2_{w_b}$. It is assumed that all channels are mutually independent. Similarly, the capacity at Eve can be expressed as
\begin{equation}
\begin{aligned} \label{eq:EveCapacity}
    C_e  =\log _2\left(\frac{|y_{eve}|^2}{\beta_5^2 + \beta_6^2+\sigma^2_{w_e}}\right),
\end{aligned}
\end{equation}
where $|y_{eve}|^2 = \beta_5^2 + \beta_6^2 + \beta_7^2 + \beta_8^2+\sigma^2_{w_e}$. The SC of the considered network, defined as the difference between the channel capacities of Bob and Eve, is given by
\begin{equation}
C_s  =\left[C_b-C_e\right]^+,
\label{eq:CS_end}
\end{equation}
where $[ \cdot ]^+ = \max(\cdot,0)$ denotes the non-negative operator. 

\vspace{-0.01cm}

\subsection{SC Maximization}
The SC can be improved by either increasing the capacity of Bob through adjusting the phase shifts of the RIS for the CS or degrading the capacity of Eve through the phase shift optimization for both CS and AN. However, higher SC may not be sufficient to guarantee confidentiality, since Eve may decode the intercepted signal if her channel quality is adequate for supporting data reception. A robust PLS system must not only aim to maximize the SC but also explicitly constrain the information rate that can be achieved by Eve. To ensure reliable secrecy, this section formulates an optimization problem that seeks to maximize the SC while keeping Eve’s capacity below a specified threshold. The optimization problem is formulated as follows
\begin{equation}
\mathbf{P:}\; \max_{\alpha_1, \alpha_2} C_s(\alpha_1, \alpha_2) 
\;\;\text{s.t.}\;\;
\left\{
\begin{aligned}
\text{C}_1: & \quad C_b \ge \bar{C}_b, \\
\text{C}_2: & \quad C_e \le \bar{C}_e, \\
\text{C}_3: & \quad \alpha_1 + \alpha_2 = 1, \\
\text{C}_4: & \quad \alpha_1, \alpha_2 \ge 0.
\end{aligned}
\right.
\end{equation}
where $\bar{C}_b = \log _2(1 + \bar{\gamma}_b )$ represents the administratively defined minimum capacity requirement for Bob, while $\bar{C}_e = \log _2(1 + \bar{\gamma}_e ) $ denotes the channel capacity limitation for Eve such that her capacity should be lower than $\bar{C}_e$. Note that $\bar{\gamma}_b$ is the minimum signal to interference noise ratio (SINR) threshold to obtain $\bar{C}_b$ and $\bar{\gamma}_e$ is the maximum SINR threshold to obtain $\bar{C}_e$. The constraint $\mathrm{C_1}$ constitutes the QoS requirement for Bob while $\mathrm{C_2}$ ensures that Eve's communication is suppressed. $\mathrm{C}_3$ enforces the total power budget, and $\mathrm{C}_4$ restricts the power allocation factors to a feasible range. Therefore, the SINR of Bob must be {\color{black} lower-bounded} as
{\small
\begin{equation}
        % \bar{\gamma}_b &\leq\frac{({ \beta_1^2 + \beta_2^2})}{({\beta_3^2 + \beta_4^2}) + {\sigma^2_{w_b}}} ,
        % \\
        \frac{\overbrace{  \alpha_1 P_{t}m_{cs}^2(  L_{sr_b} L_{r_bb}G_{sr_bb}^2 + L_{sr_e} L_{r_eb}G_{sr_eb}^2}^{({ \beta_1^2 + \beta_2^2})})}{\underbrace{\alpha_2 P_{t}m_{an}^2(  L_{ar_e} L_{r_eb}G_{ar_eb}^2 +  L_{ar_b} L_{r_bb}G_{ar_bb}^2}_{({\beta_3^2 + \beta_4^2})}) + {\sigma^2_{w_b}}} \geq \bar{\gamma}_b .
\end{equation}
}
Similarly, the SINR of Eve must be {\color{black} upper-bounded} as
{\small
\begin{equation}
        % \bar{\gamma}_e &\geq\frac{({\beta_5^2 + \beta_6^2})}{({\beta_7^2 +\beta_8^2}) + {\sigma^2_{w_e}}}  ,
        % \\ 
        % \bar{\gamma}_e \geq\frac{\overbrace{  \alpha_2 P_{t}m_{an}^2(l_{ar_e} l_{r_ee}G_{ar_ee}^2 + l_{ar_b} l_{r_be}G_{ar_be}^2}^{({\beta_5^2 + \beta_6^2})})}{\underbrace{\alpha_1 P_{t}m_{cs}^2 (  l_{sr_b} l_{r_be}G_{ar_be}^2 + l_{sr_e} l_{r_ee}G_{sr_ee}^2}_{({\beta_7^2 +\beta_8^2})}) + {\sigma^2_{w_e}}} .
        \frac{\overbrace{\alpha_1 P_{t}m_{cs}^2 (  L_{sr_b} L_{r_be}G_{sr_be}^2 + L_{sr_e} L_{r_ee}G_{sr_ee}^2}^{({\beta_7^2 +\beta_8^2})}) }{\underbrace{  \alpha_2 P_{t}m_{an}^2(L_{ar_e} L_{r_ee}G_{ar_ee}^2 + L_{ar_b} L_{r_be}G_{ar_be}^2}_{({\beta_5^2 + \beta_6^2})}) + {\sigma^2_{w_e}}}\leq \bar{\gamma}_e .
\end{equation}
}

\subsection{Practical RIS Optimization}
It is assumed that the RIS elements are restricted to discrete phase shift levels such as $0^\circ$ and $180^\circ$ for practical deployment and only the received signal power measurements are available to the optimization algorithms. Under these constraints, two practical RIS configuration methods, namely, iterative and DFT-based, are employed to determine the optimized phase shifts. Note that the RIS is partitioned into two regions: one is dedicated to enhancing Bob's capacity, and the other ensures that Eve's channel is not usable through steering the AN.

% Three figures forced to the top of the page
\begin{figure*}[!t]
  \centering
  \begin{subfigure}[t]{0.32\textwidth}
    \centering
    \includegraphics[width=\linewidth]{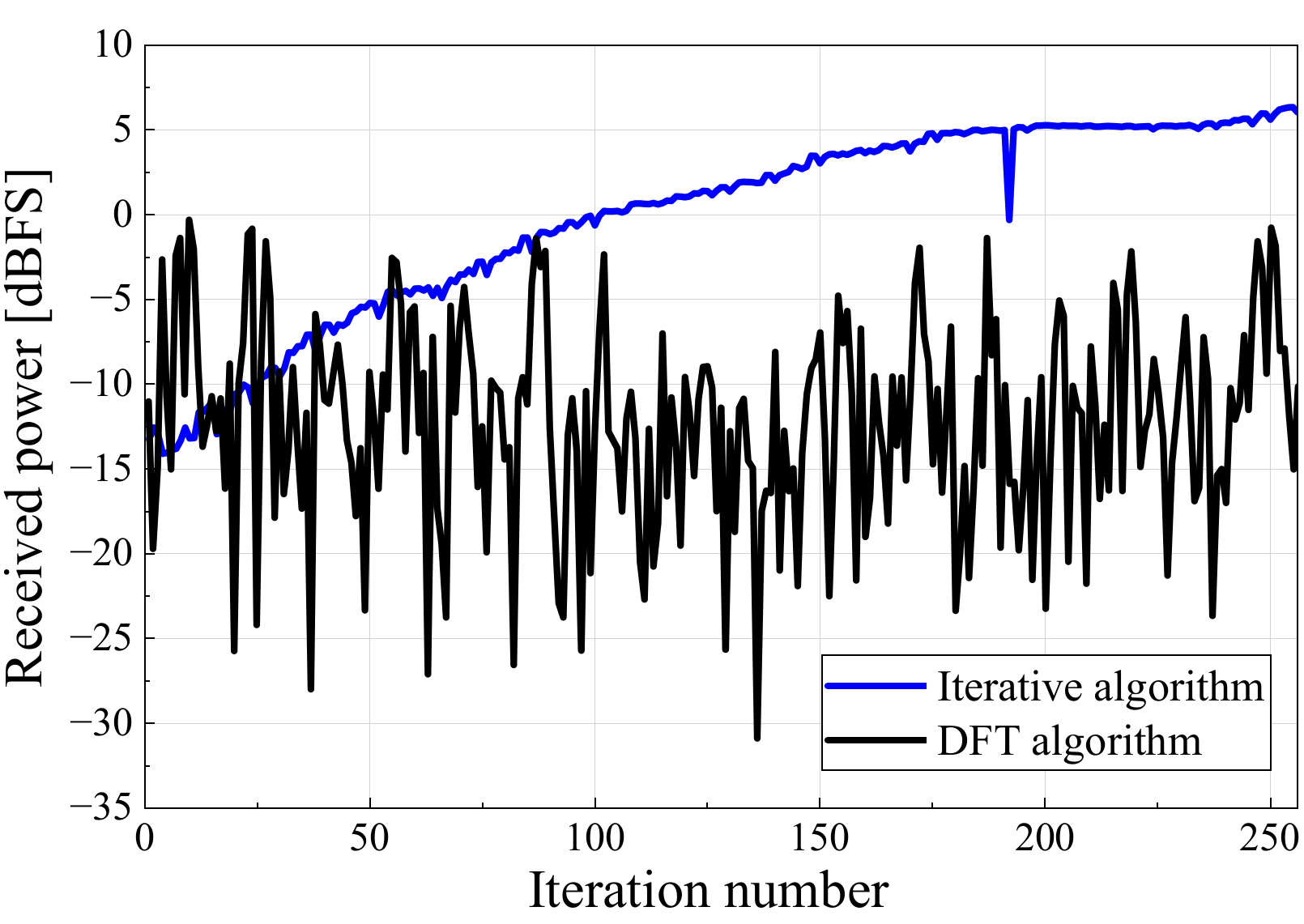} % or .pdf/.jpg
    \caption{}
    \label{fig:ue_system_combined}
  \end{subfigure}\hfill
  \begin{subfigure}[t]{0.32\textwidth}
    \centering
    \includegraphics[width=\linewidth]{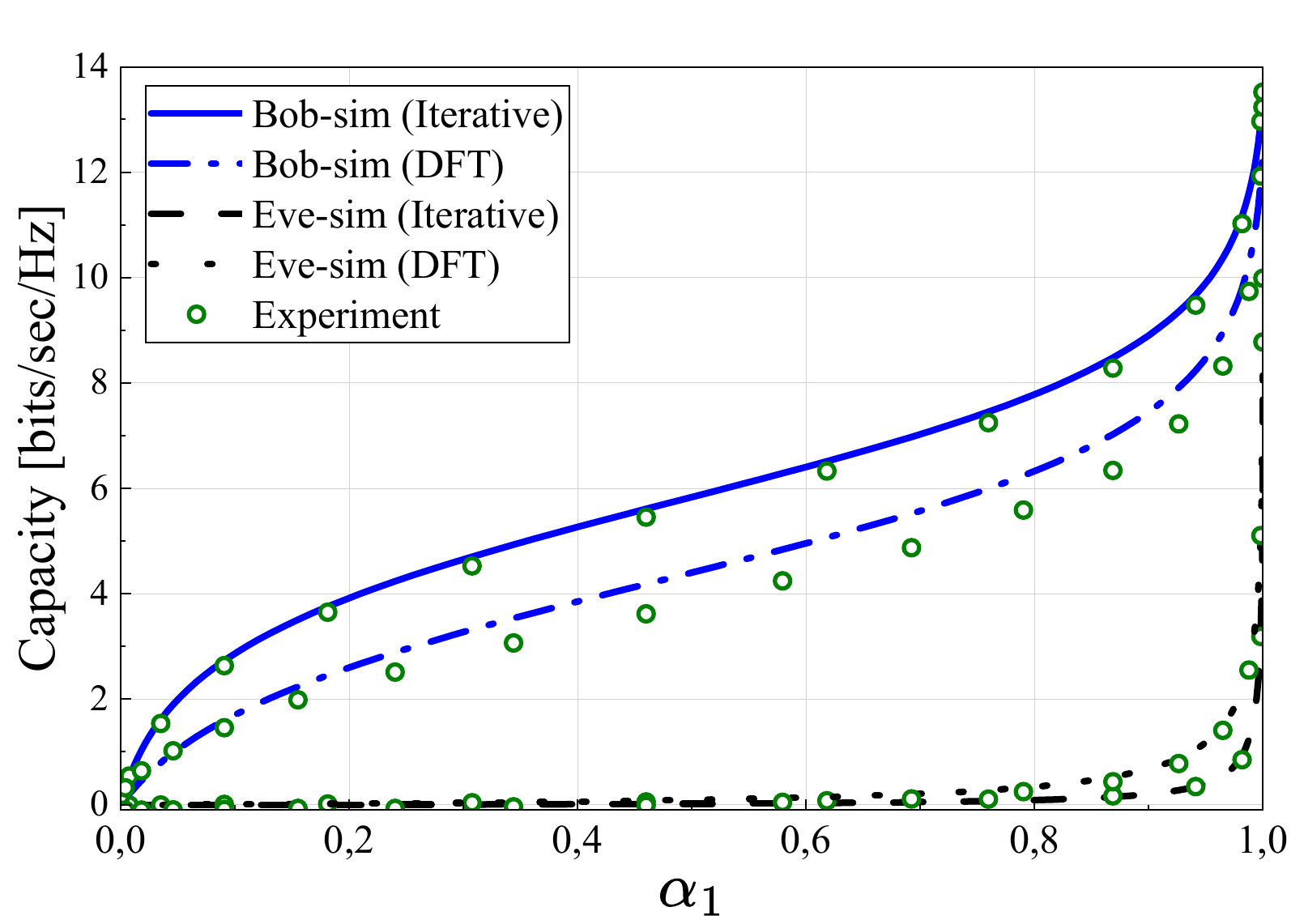}
    \caption{}
    \label{fig:C_all}
  \end{subfigure}\hfill
  \begin{subfigure}[t]{0.32\textwidth}
    \centering
    \includegraphics[width=\linewidth]{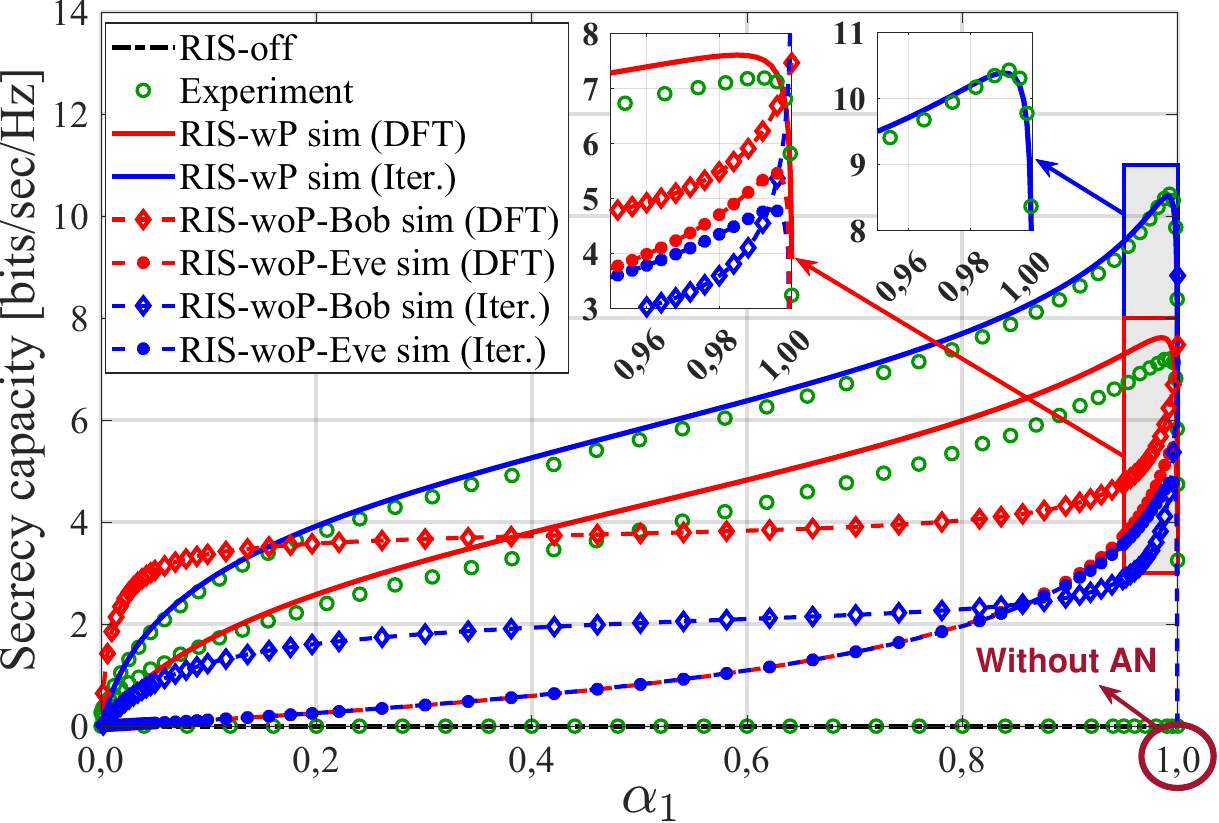}
    \caption{}
    \label{fig:SC_all}
  \end{subfigure}
  \caption{a) Bob’s received power increment through iterations; b) Capacity comparison of Bob and Eve; {\color{black}c) SC comparison}. } 
  \label{fig:three-top}\vspace{-10pt}
\end{figure*}

\subsubsection{Iterative Algorithm}
The RIS phase shift configuration is optimized using a sequential iterative approach that aims to enhance the received signal power at each iteration \cite{10423876}. Initially, $\mathbf{r}_b$ and $\mathbf{r}_e$ are set to $0^\circ$. The iterative algorithm applies two distinct phase shifts ($0^\circ$ or $180^\circ$) to each element, which is chosen randomly. The corresponding element of \textcolor{black}{$\mathbf{r}_e$} is updated using the phase shift of the selected element that maximizes the received power of the \textcolor{black}{AN} at \textcolor{black}{Eve} (i.e., \textcolor{black}{$P_{AN}$}). This procedure is iteratively applied to each subsequent element in \textcolor{black}{$\mathbf{r}_e$}. Similarly, the elements of \textcolor{black}{$\mathbf{r}_b$} are configured using this iterative approach; however, the objective is to maximize the received power of the \textcolor{black}{CS} signal at \textcolor{black}{Bob} (i.e., $\textcolor{black}{P_{CS}}$), thereby reducing her channel capacity, as given in (\ref{eq:EveCapacity}). {\color{black} Quantitatively, this criterion is defined as selecting the binary phase shift for each element in $\mathbf{r}_e$ that yields the maximum measured baseband power of the AN signal at Eve's location.} The final RIS configuration is obtained as the concatenation of $\mathbf{r}_b$ and $\mathbf{r}_e$. {\color{black} The pseudocode for the iterative approach is provided in Algorithm \ref{alg:iterative}, where $N$ and $S$ denote the number of RIS elements and the number of discrete phase shift states per element, respectively.}

\begin{algorithm}[t]\color{black} \small
\caption{\color{black} Iterative Algorithm} \label{alg:iterative}
\begin{algorithmic}[1]
\renewcommand{\algorithmicrequire}{\textbf{Input:}}
\renewcommand{\algorithmicensure}{\textbf{Output:}}
\REQUIRE $P_{CS}$, $P_{AN}$, $N$, and $S$
\ENSURE  $\mathbf{r}_b$ and $\mathbf{r}_e$
\STATE Initialize $\mathbf{r}_b$  and $\mathbf{r}_e$ with $0^\circ$ phase shifts
\STATE $P_{AN,max}$ $\gets 0$ and $P_{CS,max}$ $\gets 0$
\FOR {$i = 1$ to $N/2$}
\FOR {$j = 1$ to $S$}
\STATE Apply configuration $j$ to $i$th element
\STATE Measure the received power $P_{AN}$
\IF {($P_{AN}>P_{AN,max}$)}
\STATE update $\mathbf{r}_e$ and $P_{AN,max}$ $\gets$ $P_{AN}$
\ENDIF
\ENDFOR
\ENDFOR
\FOR {$i = N/2 +1$ to $N$}
\FOR {$j = 1$ to $S$}
\STATE Apply configuration $j$ to $i$th element
\STATE Measure the received power $P_{CS}$
\IF {($P_{CS}>P_{CS,max}$)}
\STATE update $\mathbf{r}_b$ and $P_{CS,max}$ $\gets$ $P_{CS}$
\ENDIF
\ENDFOR
\ENDFOR
\RETURN $\mathbf{r}_b$ and $\mathbf{r}_e$
\end{algorithmic}
\end{algorithm}

\subsubsection{DFT-based Codebook Algorithm}
The RIS phase shift configuration employs a pre-computed DFT-based beam codebook comprising $N$ mutually orthogonal codewords, each of which has a dimension of $N/2 \times 1$. Initially, \textcolor{black}{$\mathbf{r}_b$} is set to $0^\circ$, while the codewords are sequentially applied to \textcolor{black}{$\mathbf{r}_e$} to identify the one that maximizes \textcolor{black}{$P_{AN}$}. Conversely, by setting \textcolor{black}{$\mathbf{r}_e$} to $0^\circ$, the codewords are sequentially uploaded to \textcolor{black}{$\mathbf{r}_b$} to select the codeword that maximizes \textcolor{black}{$P_{CS}$}. The overall RIS configuration is obtained by concatenating the individual configurations of $\mathbf{r}_b$ and $\mathbf{r}_e$. {\color{black} The pseudocode for the DFT-based approach is provided in Algorithm \ref{alg:DFT}, where $\mathbf{c}_k$ is the $k$th column of the DFT matrix.}

\begin{algorithm}[t] \color{black} \small
\caption{\color{black} DFT-based Codebook Algorithm} \label{alg:DFT}
\begin{algorithmic}[1]
\renewcommand{\algorithmicrequire}{\textbf{Input:}}
\renewcommand{\algorithmicensure}{\textbf{Output:}}
\REQUIRE $P_{CS}$, $P_{AN}$, and $N$
\ENSURE  $\mathbf{r}_e$ and $\mathbf{r}_b$
\STATE Initialize $\mathbf{r}_b$  and $\mathbf{r}_e$ with $0^\circ$ phase shifts
\STATE Generate $(N/2 \times N)$ DFT matrix, where $k$th column is $\mathbf{c}_k$
\FOR {$i = 1$ to $N$}
\STATE Apply configuration codeword with $\mathbf{c}_i$ to $\mathbf{r}_e$
\STATE Measure and store the received power $P_{AN}$
\ENDFOR
\STATE $\mathbf{r}_e \gets \mathbf{c}^*$, where $\mathbf{c}^*$ is the codeword that maximized $P_{AN}$
\FOR {$j = 1$ to $N$}
\STATE Apply configuration codeword with $\mathbf{c}_j$ to $\mathbf{r}_b$
\STATE Measure and store the received power $P_{CS}$
\ENDFOR
\STATE $\mathbf{r}_b \gets \mathbf{c}^*$, where $\mathbf{c}^*$ is the codeword that maximized $P_{CS}$
\RETURN $\mathbf{r}_b$ and $\mathbf{r}_e$

\end{algorithmic}
\end{algorithm}

\textcolor{black}{The proposed iterative algorithm has complexity $\mathcal{O}(NS)$, where $N$ denotes the number of RIS elements and $S$ is the number of discrete phase shifts (e.g., $S=2$ for 1-bit resolution). The DFT-based algorithm utilizes a predefined codebook of size $C$, which is generated offline, and therefore, its online search complexity is $\mathcal{O}(C)$. As illustrated in Fig. \ref{fig:ue_system_combined}, both algorithms are executed for 256 iterations over Bob’s and Eve’s partitions, respectively, to ensure a fair comparison. }

\vspace{-13pt}
\section{Simulation and Experimental Results}
This section presents the performance evaluation of the proposed AN-based RIS-assisted PLS system through both simulation and SDR-based measurement experiments. Specifically, the RIS simulation experiments are performed in MATLAB using the \textit{helperRISsurface} function, where the RIS is divided into two parts. The right part with 128 antenna elements in a $16 \times 8$ grid is dedicated to Bob while the left part with the same number of elements is dedicated to Eve. The simulation parameters are listed in Table~\ref{tab:3dpositions}.

\begin{table}[t]
\centering
\caption{Simulation parameters}
\label{tab:3dpositions}
\begin{tabular}{p{4cm}| p{4cm}}
\toprule
\textbf{Parameter} & \textbf{Description} \\
\midrule
Carrier frequency ($f_c$) & $3.75 \ $GHz
\\
Sampling rate ($f_s$)& $0.5 \ $MHz 
\\
Transmit power ($P_t$)& $-9\ $dBm
\\
Transmit (horn) antenna gain & $13\ $dBi
\\
Noise power ($\sigma^2_{w_b},\sigma^2_{w_e}$)& $-90\ $dBm 
\\
Number of RIS elements ($N$)& $256 \ (16\times16)$
\\
RIS element spacing ($d$)& $4.1\ $cm
\\
Transmit antenna type& Cosine antenna element
\\
RIS cell antenna type& Cosine antenna element
\\
Channel model& Free space path loss
\\ 
CS transmitter location ($x,y,z$)& $\left(0.74\,,0.31\,,0\right)\,\text{m}$ 
\\
AN transmitter location ($x,y,z$)& $\left(0.74\,,-0.31\,,0\right)\,\text{m}$ \\ 
RIS location ($x,y,z$)& $\left(0,0,0.4\right)\,\text{m}$ 
\\ Bob location ($x,y,z$)& $\left(1.19\,,1.41\,, 0\right)\,\text{m}$ 
\\ Eve location ($x,y,z$)& $\left(1.19\,,-1.41\,, 0\right)\,\text{m}$ \\
\bottomrule
\end{tabular}\vspace{-11pt}
\end{table}

The measurement setup is illustrated in Fig.~\ref{fig:measurementsetup}, comprising three SDR devices {\color{black} (USRP B210)}, two sub-6~GHz horn antennas, a 256-element RIS structure with four tiles ($8 \times 8$) as described in \cite{yerliRIS}. One SDR serves as the transmitter (Alice), sending the CS and the AN simultaneously through the ports A and B, respectively, using two separate horn antennas. Both signals are quadrature phase shift keying modulated single-carrier waveforms, where the CS carries message bits, while the AN carries randomly generated bits. {\color{black} Frequency and frame synchronizations between the SDR nodes are achieved through a Barker code at the beginning as a preamble. To mitigate instantaneous noise variances during the practical optimization stage, the received power measurements for each tested RIS configuration are averaged over $10$ consecutive samples. Furthermore, the feedback and processing delays between the host PC and the SDRs are strictly bounded, ensuring the RIS configuration updates are applied well within the static channel's coherence time.} Similar to the simulation experiments, the RIS prototype is logically split into two equal segments: the right segment ($16 \times 8$) directs the CS to Bob, and the left segment directs the AN to Eve. {\color{black}In our proposed design, cross-zone interference is minimized by the spatial separation of the CS to Bob through the RIS allocated for Bob and AN to Eve through the RIS allocated for Eve paths, where the algorithms maximize received powers to steer main beams and reduce leakage. Simulations and measurements confirm that modeled sidelobe interference remains negligible with minimal impact on secrecy performance.} {\color{black}In practical deployments, RIS systems may experience coverage blind zones. Angular blind zones occur when reflection efficiency degrades at extreme angles (e.g., beyond $[-45^\circ, +45^\circ]$ for our prototype). Additionally, power blind zones can arise from severe cascaded path loss over extended distances. While our experimental setup is constrained to indoor distances and effective angular ranges to mitigate these effects, addressing these blind zones in large-scale deployments remains a critical direction for future research.
}

Figure~\ref{fig:ue_system_combined} shows the received power of the CS at Bob as a function of the iteration number for the iterative and DFT-based algorithms, each evaluated over $256$ trials. In the iterative routine, the phases of the $128$ RIS elements are toggled between $\{0^{\circ}, 180^{\circ}\}$. In contrast, the DFT-based routine sweeps $256$ orthogonal phase configurations. The iterative curve increases steadily with the iteration count, while the DFT sweep produces a more irregular profile due to the orthogonal phase patterns. When selecting the best configuration, the iterative method typically attains its maximum received power toward the end of iterations, whereas the DFT-based method reaches its peak earlier.
Using the same number of iterations, the iterative approach consistently increases $P_{CS}$ and ultimately reaches the maximum value which is higher than the DFT baseline. Therefore, when the objective is to maximize received power within a fixed number of tests, the iterative method is preferable: the DFT sweep can identify a good configuration quickly, but its best $P_{CS}$ remains lower than that attained by the iterative strategy.

Bob’s and Eve’s channel capacities using the iterative and DFT-based methods are shown in Fig.~\ref{fig:C_all}, where both simulation and experiment results for each $\alpha_{1}$ are obtained using (\ref{eq:BobCapacity}) and (\ref{eq:EveCapacity}), respectively. 
Across the range of the $\alpha_{1}$ values, Bob’s curve with the iterative method consistently remains above the DFT-based method. 
For example, at $\alpha_{1}\!\approx\!0.2$, the capacities of the iterative and DFT-based methods are about $3.9$ and $2.6$ bits/s/Hz, respectively. 
Eve’s capacity is essentially zero for $\alpha_{1}\le 0.6$ and increases when $\alpha_{1}$ gets higher. Eve's capacity increases exponentially when $\alpha_1$ gets beyond $0.8$ and reaches $\sim1$–$3$ bits/s/Hz at near $\alpha_{1}\!\approx\!0.98$, where the DFT-based method provides slightly higher results because nearly all power is devoted to data and little remains for AN. 
Overall, the iterative method provides a consistently higher capacity for Bob over the DFT sweep across the practical region $\alpha_{1}\in[0.2,0.9]$ while keeping Eve suppressed. 
The close agreement between simulations and measurements further supports the validity of the proposed PLS system.

\begin{figure}[t]
    \centering
    \includegraphics[width=0.85\linewidth]{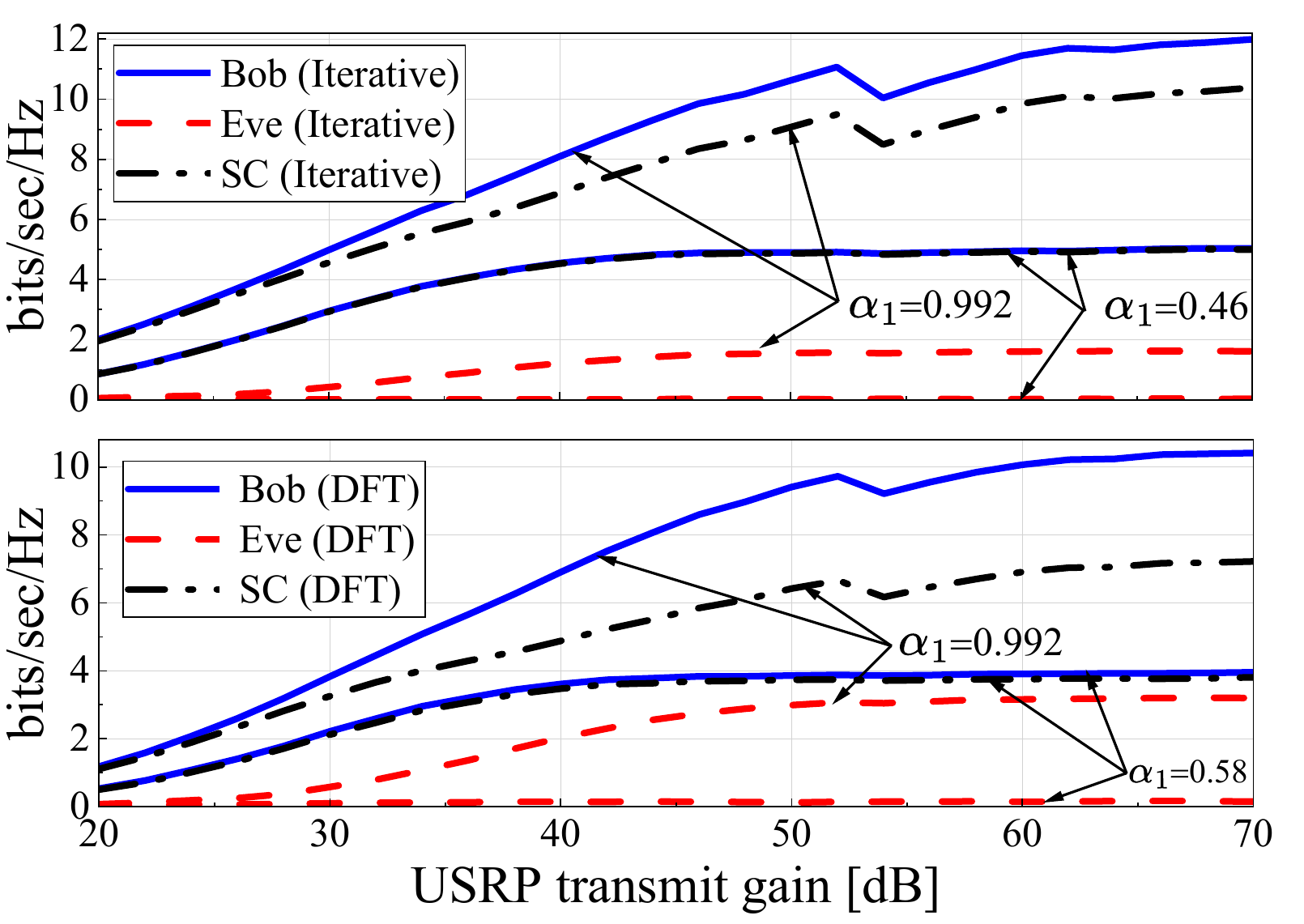}
    \caption{Experimental results of Eve's and Bob's capacities and the corresponding SC.\vspace{-0.3cm}}
    \label{fig:ITR_txpow}\vspace{-5pt}
\end{figure}

Figure~\ref{fig:SC_all} shows the SC results for the iterative, DFT-based with RIS partitioning {\color{black}(\textit{wP})} and {\color{black}\textit{RIS-off}} cases with respect to the power scaling parameter $\alpha_1$. \textcolor{black}{When $\alpha_1 = 1$, only the CS is transmitted without any AN. This \textit{Without AN} ($\alpha_1 = 1$) case is highlighted with a circle on the \textit{x}-axis in Fig. \ref{fig:SC_all}. When $\alpha_1 = 1$, the secrecy capacity drops significantly, showing that CS transmission without AN is ineffective for secure communications.}
\textcolor{black}{In the \textit{RIS-off} case, all RIS elements are assigned a phase shift of $0^\circ$, causing the RIS to act as a passive, mirror-like reflector without any intelligent phase optimization.} \textcolor{black}{We performed additional simulations, where all 256 RIS elements are dedicated to Bob (\textit{RIS-woP-Bob sim}, without RIS partitioning assigned for Bob) or Eve (\textit{RIS-woP-Eve sim}, without RIS partitioning assigned for Eve). Both configurations are tested using the iterative and DFT-based algorithms. For the DFT approach, the codebook is regenerated again to produce a 256-element configuration. The results show that for the scenarios without RIS partitioning, the DFT-based algorithm achieves higher SC values than the iterative method.} The maximum SC is achieved around $\alpha_1 = 0.992$ for both iterative and DFT algorithms. However, Fig.~\ref{fig:C_all} shows that the capacity of Eve is relatively high when $\alpha_1$ is beyond $0.6$. This may pose a potential security risk as Eve may still decode the CS. Therefore, the scaling factor can be selected to minimize this risk at the expense of lower SC, which is shown in Fig.~\ref{fig:ITR_txpow}. 

Bob’s and Eve’s channel capacities and the corresponding SC results are depicted in Fig.~\ref{fig:ITR_txpow} as a function of the SDR transmit gain \cite{EttusB200_RF_Performance}. The $\alpha_1$ value of $0.992$ corresponds to the maximum SC when there is no constraint on Eve's capacity. As shown in Fig.~\ref{fig:C_all}, $\alpha_1$ is determined as $0.46$ and $0.58$ for the iterative and DFT-based methods, respectively, so that Eve's capacity is lower than $1\%$ of Bob's capacity. The results across all SDR transmit gains show that Eve's capacity is successfully suppressed to provide secure communication when the $\alpha_1$ values are $0.46$ and $0.58$ for the iterative and DFT-based methods, respectively. Note that, for all SDR transmit gains, the iterative algorithm consistently achieves a higher SC. 

\begin{figure}[t]
    \centering
    \includegraphics[width=0.9\linewidth]{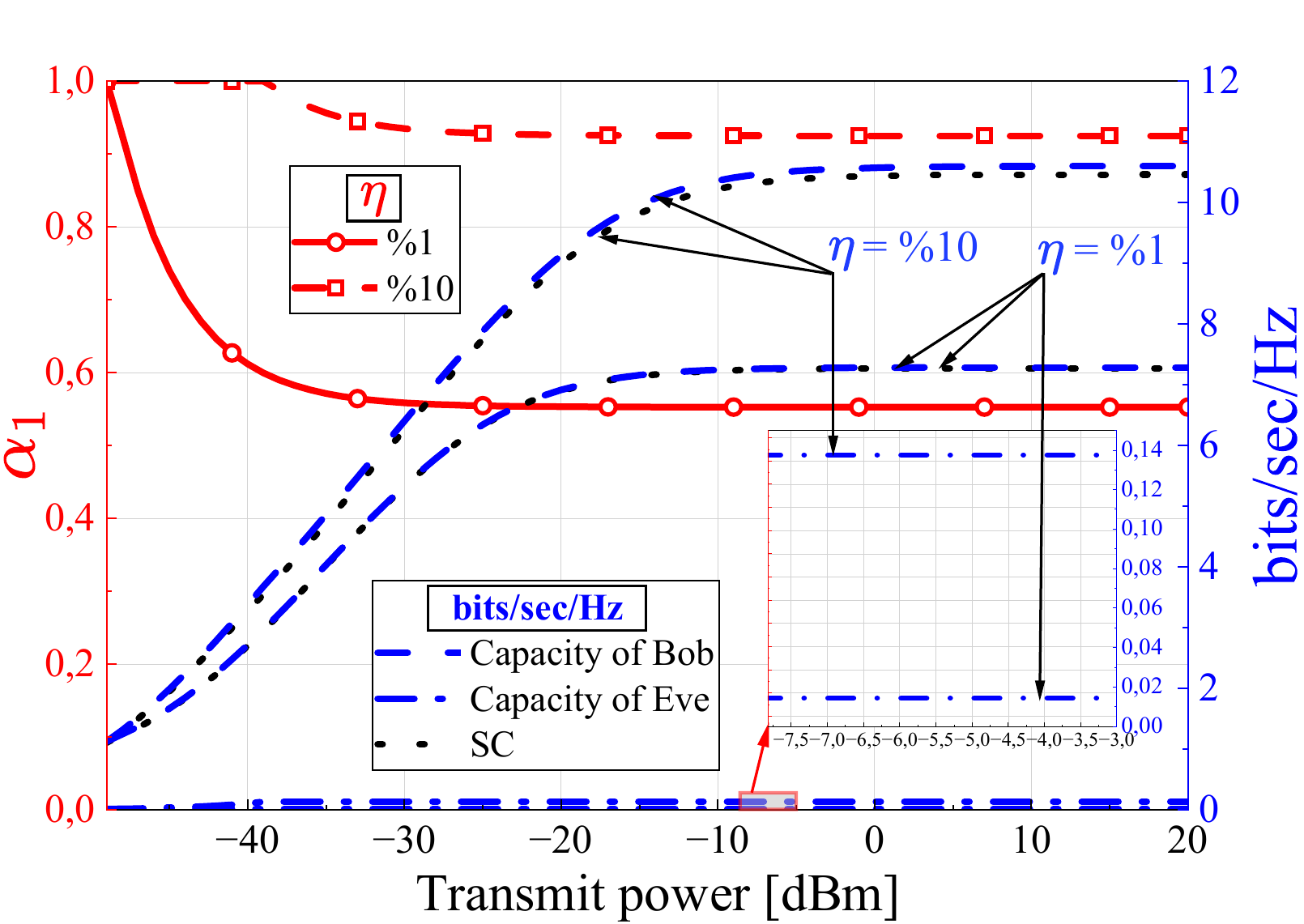}
    \caption{CVX-based analysis of Alice's transmit power impact on $\alpha_1$ and the capacity values under different QoS ratios.}
    \label{fig:ITRCVX_all}\vspace{-3pt}
\end{figure}

Figure \ref{fig:ITRCVX_all} presents the simulation results for Bob’s and Eve’s channel capacities, and the corresponding SC values as a function of the transmit power for the optimum values of $\alpha_1$ that are calculated using the Matlab CVX tool when $\eta$ is set to $1\%$ and $10\%$, where $\eta = \frac{\bar{\gamma_e}}{\bar{\gamma_b}}$ represents the ratio of Eve's and Bob's SINR constraints. Then, the iterative algorithm using the optimum $\alpha_1$ values is employed to obtain the channel capacities and the corresponding SC results. On the left $y$ axis, the values of $\alpha_1$ converge to $\alpha_1 \sim 0.55$ and $\alpha_1 \sim 0.92$ corresponding to $\eta = 1\%$ and $\eta = 10\%$, respectively, when the total transmit power increases. In the right $y$ axis, Bob's capacities converge to slightly higher than $7$ bits/s/Hz and slightly lower than $11$ bits/s/Hz for $\eta = 1\%$ and $\eta = 10\%$, respectively. When $\eta$ decreases from $10\%$ to $1\%$, the optimal value of $\alpha_1$ decreases from $\alpha_1 \sim 0.92$ to $\alpha_1 \sim 0.55$. Figure \ref{fig:ITRCVX_all} presents the simulation results for Bob’s and Eve’s channel capacities, and the corresponding SC values as a function of the transmit power for the optimum values of $\alpha_1$ that are calculated using the Matlab CVX tool when $\eta$ is set to $1\%$ and $10\%$, where $\eta = \frac{\bar{\gamma_e}}{\bar{\gamma_b}}$ represents the ratio of Eve's and Bob's SINR constraints. Then, the iterative algorithm using the optimum $\alpha_1$ values is employed to obtain the channel capacities and the corresponding SC results. On the left $y$ axis, the values of $\alpha_1$ converge to $\alpha_1 \sim 0.55$ and $\alpha_1 \sim 0.92$ corresponding to $\eta = 1\%$ and $\eta = 10\%$, respectively, when the total transmit power increases. In the right $y$ axis, Bob's capacities converge to slightly higher than $7$ bits/s/Hz and slightly lower than $11$ bits/s/Hz for $\eta = 1\%$ and $\eta = 10\%$, respectively. When $\eta$ decreases from $10\%$ to $1\%$, the optimal value of $\alpha_1$ decreases from $\alpha_1 \sim 0.92$ to $\alpha_1 \sim 0.55$. This is due to the fact that higher transmit power should be allocated to AN to effectively suppress Eve’s SINR. However, this results in lower SC values. These results show that our proposed AN-driven RIS-assisted PLS system can effectively configure the power scaling parameter to balance the trade-off between Bob's and Eve's channel capacities and the corresponding the SC results. These results show that our proposed AN-driven RIS-assisted PLS system can configure the power scaling parameter to balance the trade-off between Bob's and Eve's capacities and the corresponding SC results. 

\section{Conclusion}
This paper proposed and experimentally validated an AN-driven RIS-assisted PLS scheme. The RIS was divided into two equal segments: one to enhance the received CS power at Bob and the other to increase interference at Eve. Two practical RIS optimization methods, iterative and DFT-based, were implemented to configure the RIS phase shifts, along with a power allocation strategy to balance the CS and AN signals under different secrecy requirements. Simulation and experimental results showed notable SC improvements, with the iterative method outperforming the DFT-based approach when the number of iterations equals the number of DFT codewords. These results confirm the feasibility of jointly using RIS and AN to enhance PLS in practical wireless systems. Future work will investigate dynamic RIS partitioning, multiple eavesdroppers and legitimate users, and multi-phase adjustment strategies.

%In this paper, an AN-driven RIS-assisted PLS scheme is proposed and experimentally validated. The RIS is partitioned into two equal segments, where one is utilized to enhance the received CS power at Bob, and the other is used to increase the interference at Eve. Two practical RIS optimization methods, namely iterative and DFT-based, are implemented to configure the phase shifts of the RIS antenna elements. Additionally, a power allocation strategy is introduced to optimize the transmission power between the CS and the AN signal, enabling adaptability to different secrecy requirements. Simulation and experimental results demonstrate that the proposed approach provides notable SC improvements. In particular, the iterative method is shown to outperform the DFT-based approach in terms of the SC when the number of iterations in the iterative method is equal to the number of codewords in the DFT-based method. These findings confirm the feasibility of employing the RIS and the AN jointly to enhance PLS in practical wireless communication systems. In future work, this study will be extended to comprehensively evaluate the impact of dynamic RIS partition sizes to further maximize performance, as well as to secure communications against multiple eavesdroppers and legitimate users with a multi-phase adjustment.
\section*{Acknowledgment}
\fontdimen2\font=0.54ex
This research was supported by Nazarbayev University under the Faculty Development Competitive Research Grants program \textnumero 110326FD3228 (PI: S.A.) and  \textnumero  110326FDCRGP3216 (PI: G.N.).

\balance
\bibliographystyle{IEEEtran}

\vspace{-0.2cm}
\bibliography{references.bib}

\end{document}